# THE EFFICIENCY OF QUANTUM ENGINES USING THE PÖSCHL – TELLER LIKE OSCILLATOR MODEL.

**E. O. Oladimeji[1]**

Department of Physical Sciences, Joseph Ayo Babalola University (JABU)
Ikeji-Arakeji, Nigeria.

## ABSTRACT

In this work, the quantum-mechanical properties of the strongly non-linear quantum oscillator described by the Pöschl-Teller [PT] model is examined. This model has a relation with two well-known models, the free particle [FP] in a box and the harmonic oscillator [HO], as described in our previous works [1–3]. Using the [PT] model, a quantum-mechanical analog of the Joule-Brayton [JB] cycle and Otto cycle have been constructed through changes of both, the width of the well and its the quantum state. The efficiency of quantum engines based on the Pöschl-Teller-like potential is derived, which is analogous to classical thermodynamic engines.

## 1. INTRODUCTION

Heat engines have been a subject of interest and intense research since the eighteenth century due to their practical applications. Of recent, diverse efforts have been made to understand the working mechanism of heat engines [4–7] which has led to the introduction of several Quantum analogues [8–11]. Some experiments have led to the miniaturization of thermal engines down to microscale [12–16], where the working engine is a quantum system which can be termed technically as Quantum heat engines [QHE]. This field considers analogies between quantum systems and macroscopic engines, with an example in the proposed model by *Bender et al* [17] of a cyclic engine based on a single quantum mechanical particle of mass $m$ confined in an infinite one-dimensional potential well of width $L$ (free particle [FP] in the box). This model replaces the role of piston in a cylinder and temperature in classical thermodynamics to the walls of the confining potential and energy as given by the pure-state expectation value of the Hamiltonian in Quantum thermodynamics respectively.

Following the formulation of *Bender et.al.* [17], we replace the concept of temperature with the expectation value of the Hamiltonian, that is, the ensemble average of the energies of the quantum particle. We also replace the classical thermodynamic idea of a large number of gas molecules trapped in a volume by an infinite number of copies of a quantum particle trapped in a potential well. Thus, what in a classical system is described as "an ensemble in contact with a heat bath where the walls of the potential move", here implies keeping the quantum ensemble, by some unspecified physical means, with a constant expectation value of the Hamiltonian while the classical parameter that describes the potential width moves.

In this paper, we introduced the Pöschl-Teller [PT] model, where a single quantum-mechanical particle is confined in the Pöschl-Teller oscillator [PTO]. This model introduced a family of anharmonic [PT] − Potentials $V(x)$ that allows the exact solutions of the one-dimensional Schrödinger equation in the coordinate, or *x-representation* [18,19]. The exact solution in simple term is the combination of both the Bloch's harmonic oscillator [HO], which unconfined in space and the quasi-free particle [FP] in the box with impenetrable hard walls where $\lambda(L) \gg 1$ $and$ $0$ respectively [1,20]. This implies that when $\lambda(L) \gg 1$ the harmonic term dominates, while when $\lambda(L) = 0$ the particle-in-the-box term remains.

---
[1] *Corresponding author: E. O. Oladimeji. E-mail: nockjnr@gmail.com; eooladimeji@jabu.edu.ng*

$$V(x; L) = -V(x; L) = V_0 tg^2[\alpha(L)x]; \; \alpha(L) = \pi/L \tag{1}$$

At $x = \pm L$ the potential becomes singular, which physically means the presence of a pair of impenetrable walls. The Pöschl-Teller [PT] model allows the introduction of pressure operator $\hat{H}(\hat{x}, \hat{p}, L)$, which according to Hellmann and Feynman [21,22] is connected with the energy operator or the Hamiltonian $\hat{H}(\hat{x}, \hat{p}, L) = (\hat{p}^2/2m) + \hat{V}(\hat{x}; L)$ by the formal relation: $\hat{P}(\hat{x}, \hat{p}, L) = -(\partial/\partial L)H(\hat{x}, \hat{p}, L)$, where the relation between the pressure $\hat{P}$ and energy $\hat{E}$ operator is defined as:

$$\hat{P}_n(L) = -\frac{\partial \hat{E}_n}{\partial L} \tag{2}$$

This implies that the Hamiltonian $\hat{H}$ depends parametrically on $L$, therefore, it's some form of Born-Oppenheimer approximation where the degree of freedom $L$ is treated semi-classically [23,24].

The potential (1) leads to an exact solution of the Schrodinger equation with fully discrete positive energy levels $E_n(L) > 0$ (including the ground level $E_1(L)$) [1,20]:

$$E_n^{PT}(L) = W(L)[n^2 + \lambda(L)(2n + 1)] \tag{3}$$

Where; $W(L) = \pi^2 \hbar^2 / 2mL^2$ and $\lambda(L) = [(2/(\pi\zeta(L))^2 + 1]^{1/2} - 1$ as defined in [1].

Since the pressure operator $P_n(L) = (s/L)E_n(L)$, where $s = 2$ therefore;

$$P_n^{PT}(L) = \frac{2W(L)}{L}[n^2 + 2\lambda(L)(n + \tfrac{1}{2})\{1 - \mu(L)\}] \tag{4}$$

Where; $\mu(L) = 1 - [\lambda(L) - 1][2\lambda(L) - 1]^{-1}$.

## 2. THE JOULE – BRAYTON CYCLE.

The classical Joule-Brayton cycle is composed of two isobaric and two adiabatic processes *(see Fig.1)* each of which is reversible.

During the isobaric process, the pressure remains constant even when the system is compressed or expanded (i.e the rate of change of energy with respect to the change in width $L$ of the well is constant). The energy value as a function of $L$ may be written as:

$$E(L) = \sum_{n=1}^{\infty} |a_n|^2 E_n \tag{5}$$

where $E_n$ is the energy spectrum (3) and the coefficients $|a_n|^2$ are constrained by the normalization condition $\sum_{n=1}^{\infty}|a_n|^2 = 1$. In the quantum mechanical case, if we assume that the system at the initial state $\psi_n(x)$ of volume $L$ is a linear combination of eigenstates $\phi_n(x)$, the expectation value of the Hamiltonian changes with respect to the change in width $L$ of the well, then the instantaneous pressure exerted on the walls can be obtained using the relation (2).

**Process 1: Isobaric Expansion**

Given that the system expands isobarically from its initial state $n = 1$ at point 1 (i.e. from $L = L_1$ to $L = L_2$) and is excited into the second state $n = 2$, keeping the expectation value of the Hamiltonian constant. Thus, the state of the system is a linear combination of its two energy eigenstates,

$$\Psi_n = a_1(L)\phi_1(x) + a_2(L)\phi_2(x),$$

where $\phi_1$ and $\phi_2$ are the wave functions of the first and second states respectively. The coefficients satisfy the condition $|a_1|^2 + |a_2|^2 = 1$. The expectation value of the Hamiltonian in this state as a function of $L$ is calculated as $E = \langle \psi|H|\psi \rangle$:

$$E = W(L)\big[4 + 5\lambda(L) - (3 + 2\lambda(L))|a_1|^2\big], \tag{6}$$

The pressure during this process remains constant and its value is given in terms of its definition (2):

$$P = -\frac{dE}{dL} = \frac{2W(L)}{L}[4 + 5\lambda(L)\{1 - \mu(L)\} - (3 + 2\lambda(L)\{1 - \mu(L)\})|a_1|^2] \tag{7}$$

The pressure at point 1 as a function of $L_1$ is

$$P_A = P_1^{PT} = \frac{2W(L)}{L_1}; \tag{8}$$

equating (7) and (8) one can conclude

$$L = L_1[4 + 5\lambda(L)\{1 - \mu(L)\} - (3 + 2\lambda(L)\{1 - \mu(L)\})|a_1|^2]$$

Thus, the maximum possible value of $L$ during this isothermal expansion is $L = L_2$, since $a_1 = 0$ at (point 2): Therefore $L = L_1[4 + 5\lambda(L)\{1 - \mu(L)\}]$,

$$L = L_1[4 + 5\lambda(L)\{1 - \mu(L)\}] = L_2$$

By combining (6) and (7), we observe the energy $E$ as a function of $L$ to be

$$E = \frac{W(L)L}{L_1}, \tag{9}$$

or $\frac{E}{L} = \frac{W(L)}{L_1} = const$, which is analogue to *isobaric equation*.

**Process 2: Adiabatic Expansion**

Next, the system expands adiabatically from $L = L_2$ until $L = L_3$. During this expansion, the system remains in the second state $n = 2$ as no external energy comes into the system and the change in the internal energy equals to the work performed by the walls of the well. The expectation value of the Hamiltonian is:

$$E = \frac{\pi^2 \hbar^2}{2mL^2}[4 + 5\lambda(L)]$$

the pressure is given by:

$$P_2 = \frac{2W(L)}{L}[4 + 5\lambda(L)\{1 - \mu(L)\}] \tag{10}$$

The product $LP_2(L)$ in (10) is a constant what is considered as the quantum analogue of the classical *adiabatic process*.

**Process 3: Isobaric Compression**

The system is in the second state $n = 2$ at point 3 (i.e. from $L = L_3$ until $L = L_4$), and it compresses isobarically. The system is compressed back to the initial state $n = 1$ as the expectation value of the Hamiltonian remains constant. Thus, the state of the system is a linear combination of its two energy eigenstates.

$$\Psi_n = b_1(L)\phi_1(x) + b_2(L)\phi_2(x)$$

The expectation value of the Hamiltonian in this state as a function of $L$ is calculated by means of $E = \langle \psi|H|\psi \rangle$, which result in

$$E(L) = W(L)\left[1 + 3\lambda(L) + (3 + 2\lambda(L))|b_2|^2\right] \tag{11}$$

The pressure during this process remains constant and its value is given as

$$P = -\frac{dE}{dL} = \frac{2W(L)}{L}[1 + 3\lambda(L)\{1 - \mu(L)\} + (3 + 2\lambda(L)\{1 - \mu(L)\})|b_2|^2] \tag{12}$$

The pressure at point 3 as a function of $L^3$ is:

$$P_B = P_2^{PT} = \frac{4W(L)}{L_3} \tag{13}$$

Equating (12) and (13) we can conclude

$$\frac{4}{L_3} = \frac{2[1 + 3\lambda(L)\{1 - \mu(L)\} + (3 + 2\lambda(L)\{1 - \mu(L)\})|b_2|^2]}{L}$$

therefore:

$$L = L_3\left[\frac{1 + 3\lambda(L)\{1 - \mu(L)\} + (3 + 2\lambda(L)\{1 - \mu(L)\})|b_2|^2}{2}\right]$$

Thus, the maximum possible value of L in this isothermal expansion is $L = L_4$:

$$L = L_3\left[\frac{1 + 3\lambda(L)\{1 - \mu(L)\}}{2}\right] = L_4 \tag{14}$$

and this is achieved when $b_2 = 0$ at (point 4). By combining (11) and (12), we observe the energy $E$ as a function of $L$ to be:

$$E = \frac{2W(L)}{L_3}. \tag{15}$$

**Process 4: Adiabatic Compression**

The system returns in the initial state $n = 1$ at point 4 (i.e. from $L = L_4$ until $L = L_1$), as it compresses adiabatically. The expectation of the Hamiltonian is given by $E = W(L)[1 + 3\lambda(L)]$ and the pressure applied to the potential well's wall is:

$$P_4(L) = \frac{2W(L)}{L}[1 + 3\lambda(L)\{1 - \mu(L)\}] \tag{16}$$

The work $R$ performed by the quantum heat engine during one closed cycle, along the four processes described above is the area of the closed loops represented in the *Fig.1*. By using eqs. (8), (10), (13) and (16) one obtains

$$R = R_{12} + R_{23} + R_{34} + R_{41}$$

$$R = \int_{L_1}^{L_2} P_A dL + \int_{L_2}^{L_3} P_2(L) dL + \int_{L_3}^{L_4} P_B dL + \int_{L_4}^{L_1} P_4(L) dL$$

$$R = \int_{L_1}^{L_1[4+5\lambda(L)\{1-\mu(L)\}]} \frac{2W(L)}{L_1} dL + \int_{L_1[4+5\lambda(L)\{1-\mu(L)\}]}^{L_3} \frac{2W(L)}{L}[4 + 5\lambda(L)\{1 - \mu(L)\}] dL$$
$$+ \int_{L_3}^{L_3[1+3\lambda(L)\{1-\mu(L)\}/2]} \frac{4W(L)}{L_3} dL + \int_{L_3[1+3\lambda(L)\{1-\mu(L)\}/2]}^{L_1} \frac{2W(L)}{L}[1 + 3\lambda(L)\{1 - \mu(L)\}] dL$$

$$R = P_A L_1[3 + 5\lambda(L)\{1 - \mu(L)\}] + \left[\frac{2W(L)}{L_3}[4 + 5\lambda(L)\{1 - \mu(L)\}] - \frac{W(L)}{L_1}\right] + P_B L_3\left[\frac{3\lambda(L)\{1-\mu(L)\}-1}{2}\right] +$$
$$\left[\frac{2W(L)}{L_1}[1 + 3\lambda(L)\{1 - \mu(L)\}] - \frac{4W(L)}{L_3}\right] \tag{17}$$

where $P_A$ and $P_B$ are as given in (8) and (13). The heat input $Q_H$ along the first isobaric process (1→2) is the sum of the work performed $R_{12}$ and the change in the internal energy $\Delta E_{12}$ along the isobaric process i.e. $Q_H = R_{12} + \Delta E_{12}$. The change in internal energy $\Delta E_{12}$ can be derived from (9),

$$\Delta E_{12} = \int_{L_1}^{L_1[4+5\lambda(L)\{1-\mu(L)\}]} \frac{dE(L)}{dL} dL = \frac{1}{2} P_A L_1[3 + 5\lambda(L)\{1 - \mu(L)\}] \tag{18}$$

where we used (6), and the work $R_{12}$ is given by the first term of the right of (17). Thus, the heat input $|Q_H|$ can be expressed as:

$$Q_H = \frac{3}{2} P_A L_1[3 + 5\lambda(L)\{1 - \mu(L)\}] \tag{19}$$

Finally, the efficiency of the closed cycle is defined as:

$$\eta = \frac{R}{Q_H}, \tag{20}$$

where $R$ is the work and $Q_H$ is the heat input given in the (15) and (19) respectively.

$$\eta = 1 - \frac{P_B}{P_A}\left[\frac{1}{[4+5\lambda(L)\{1-\mu(L)\}]^{1/3}}\frac{L_3}{L_1}\right] \qquad (21)$$

If we take the quotient between the pressures given by the (8) and (13), we obtain $\frac{L_3}{L_1} = \left[\frac{1}{[(4-5\lambda(L))^{1/3}-3\lambda(L)-1]}\right]^{1/3}\left(\frac{P_A}{P_B}\right)^{1/3}$. Using this relation in (21) and defining the ratio $R_P = P_A/P_B$, the efficiency can be written as:

$$\eta = 1 - \frac{1}{R^{2/3}}. \qquad (22)$$

Note that this efficiency is *analogous* to that of a classical *Joule-Brayton cycle* [25].

## 3. THE OTTO CYCLE

The classical Otto cycle is composed of two isochoric and two adiabatic processes a reversible cycle process as shown in *Fig.2* each of which is reversible.

The isochoric process is one in which the volume of the potential well is constant. During this process the system can increase or diminish its energy and the force exerted on the walls also changes according with the energy pumping by an external source.

THE

**Step 1: Isochoric Expansion**
Given that the system is in the initial state $n = 1$ at point 1 (i.e. $P = P_1$), and it expands isochorically (i.e $L$ is constant ($L_1 = L_2$)). It's excited into the second state $n = 2$ at point 2 (i.e $P = P_2$) as the expectation value of the Hamiltonian is kept constant. the state of the system is a linear combination of its two energy eigenstates,

$$\Psi_n = a_1(L)\phi_1(x) + a_2(L)\phi_2(x),$$

where $\phi_1$ and $\phi_2$ are the wave functions of the first and second states respectively. The coefficients satisfy the condition $|a_1|^2 + |a_2|^2 = 1$. The expectation value of the Hamiltonian in this state as a function of $L$ is calculated as $E = \langle\psi|H|\psi\rangle$:

$$E = W(L)[4 + 5\lambda(L) - (3 + 2\lambda(L))|a_1|^2], \qquad (23)$$

The pressure during this process remains constant and its value is given in terms of definition (2):

From the values of pressure in (2) we can express the length $L$ as:

$$L^3 = \frac{2W(L)}{P}[4 + 5\lambda(L)\{1 - \mu(L)\} - (3 + 2\lambda(L)\{1 - \mu(L)\})|a_1|^2], \qquad (24)$$

The length at point 1:

$$L_1 = \frac{2W(L)}{P_1}, \tag{25}$$

Equating (23) and (24) we can conclude:

$$P = P_1[4 + 5\lambda(L)\{1 - \mu(L)\} - (3 + 2\lambda(L)\{1 - \mu(L)\})|a_1|^2] \tag{26}$$

Thus, the maximum possible value of $P$ in this isochoric expansion is $P = P_2$:

$$P = P_1[4 + 5\lambda(L)\{1 - \mu(L)\}] = P_2 \tag{27}$$

and this is achieved when $|a_1|^2 = 0$ at point 2. By combining Equation (2) and (27), we find that the energy $E$ can be written as:

$$E = \frac{W(L)}{2}\frac{P}{P_1}, \tag{28}$$

so that the relation $\frac{E}{P} = \frac{2W(L)}{2P_1}$ stays constant which is *analogous* to classical isochoric situation.

**Step 2: Adiabatic Expansion**

Next, the system expands adiabatically from $L = L_2$ until $L = L_3$. During this expansion, the system remains in the second state $n = 2$ as no external energy comes into the system and the change in the internal energy equals to the work performed by the walls of the well. The expectation value of the Hamiltonian is:

$$E(L) = E_2^{PT} = W(L)[4 + 5\lambda(L)] \tag{29}$$

and the pressure is given by:

$$P_2 = \frac{2W(L)}{L}[4 + 5\lambda(L)\{1 - \mu(L)\}] \tag{30}$$

The product $L^3 P_2(L)$ in (30) is a constant what may be considered as the quantum *analogue* of the classical adiabatic process.

**Step 3: Isochoric Compression**

The system is in the second state $n = 2$ at point 3 (i.e. $P = P_3$), and it compresses isochorically i.e. $L_3 = L_4$. The system is compressed back to the initial state $n = 1$ at point 4 (i.e. $P = P_4$). During this compression energy is being extracted so that the expectation value of the Hamiltonian remains constant. Thus, the state of the system is a linear combination of its two energy eigenstates.

$$\Psi_n = b_1(L)\phi_1(x) + b_2(L)\phi_2(x)$$

The expectation value of the Hamiltonian in this state as a function of $L$ is calculated by means of $E = \langle\psi|H|\psi\rangle$, which result in

$$E(L) = W(L)\big[1 + 3\lambda(L) + (3 + 2\lambda(L))|b_2|^2\big] \tag{31}$$

The pressure during this process remains constant and its value is given as:

$$P = -\frac{dE}{dL} = \frac{2W(L)}{L}[1 + 3\lambda(L)\{1 - \mu(L)\} + (3 + 2\lambda(L)\{1 - \mu(L)\})|b_2|^2] \tag{32}$$

Therefore, we can express the length $L$ as:

$$L = \frac{2W(L)}{P}[1 + 3\lambda(L)\{1 - \mu(L)\} + (3 + 2\lambda(L)\{1 - \mu(L)\})|b_2|^2] \tag{33}$$

The length at point 3:

$$L_3 = \frac{2W(L)}{P_3} \tag{34}$$

Equating (32) and (33) we may conclude that:

$$P = P_3[1 + 3\lambda(L)\{1 - \mu(L)\} + (3 + 2\lambda(L)\{1 - \mu(L)\})|b_2|^2] \tag{35}$$

Thus, the maximum possible value of $P$ in this isochoric expansion is $P = P_4$:

$$P = P_3[1 + 3\lambda(L)\{1 - \mu(L)\}] = P_4 \tag{36}$$

and this is achieved when $b_2 = 0$ at (point 2). By combining Equation (32) and (36), we find the energy $E$ can be written as:

$$E = 2W(L)\frac{P}{P_3}; \tag{37}$$

this expression implies that $\frac{E}{P} = \frac{2W(L)}{P_3}$ is constant in analogy with the classic isochoric equation.

**Step 4: Adiabatic Compression**

The system returns to the initial state $n = 1$ at point 4 (i.e., from $L = L_4$ until $L = L_1$), as it compresses adiabatically. During this compression, the expectation of the Hamiltonian is given by:

$$E = W(L)[1 + 3\lambda(L)], \tag{38}$$

whereas the pressure applied to the potential well's wall is:

$$P_4 = \frac{2W(L)}{L}[1 + 3\lambda(L)\{1 - \mu(L)\}]. \tag{39}$$

The work $W$ performed by the quantum heat engine during one closed cycle, along the four steps described above is the area of the closed loops represented in the *Fig. (2).* can be related to the heat exchange as:

$$R = |Q_H| - |Q_C| \tag{40}$$

where $|Q_H|$ and $|Q_C|$ are the heat input and output during the process $1 \to 2$ and $3 \to 4$, respectively. This quantity can be calculated by using (34) and (40),

$$|Q_H| = \Delta E_{12} = \int_{P_1}^{P_1[4+5\lambda(L)\{1-\mu(L)\}]} \frac{dE}{dP} dP = \frac{W(L)}{2}[3 + 5\lambda(L)\{1-\mu(L)\}] \quad (41)$$

and

$$|Q_C| = \Delta E_{34} = \int_{P_3}^{P_3[1+3\lambda(L)\{1-\mu(L)\}]} \frac{dE}{dP} dP = -2W(L)[1 - 3\lambda(L)\{1-\mu(L)\}] \quad (42)$$

Finally, the efficiency of the closed cycle is defined as:

$$\eta = 1 - \frac{|Q_c|}{|Q_H|}$$

$$\eta = 1 - \frac{L_1^2}{L_3^2}\left[\frac{[3\lambda(L)\{1-\mu(L)\} - 3]}{[5\lambda(L)\{1-\mu(L)\} - 3]}\right]$$

we have introduced the compression ratio, $R_L = L_3 / L_1$. This result is the analogue to the efficiency of the classical Otto's cycle.

$$\eta = 1 - \frac{1}{R_L}\left[\frac{[3\lambda(L)\{1-\mu(L)\}-3]}{[5\lambda(L)\{1-\mu(L)\}-3]}\right] \quad (43)$$

## 4. OUR RESULT

From the efficiency of quantum Joule-Brayton and Otto cycle which was derived using the Pöschl-Teller [PT] Model in (21) and (43) respectively, it's necessary to compare our results with earlier work of *L.Guzmán-Vargas et al* [25] that implemented the free particle [FP] in the box model where value of $\lambda(L)$ as stated in [1] must be 1 (i.e. $\lambda(L) = 1$) therefore the efficiency is:

$$\eta = 1 - \frac{1}{4^{1/3}} \frac{P_B}{P_A} \frac{L_3}{L_1} \quad (44)$$

and

$$\eta = 1 - \frac{1}{R_L} \quad (46)$$

This is remarkable, because the result of the efficiency when $\lambda(L) = 1$ is exactly the same as in [25] which is analogous to classical *Joule-Brayton* and *Otto* cycle.

## 5. CONCLUSION

This work was motivated by the consideration of the quantum-mechanical properties of the strongly non-linear quantum oscillator in Pöschl-Teller [PT] model, where the dynamic properties of our model in Joule-Brayton [JB] and Otto cycles are explored. This was implemented by using the [PT] model to analyze cycles of quantum engines such that the derived equations are analogous to classical isobaric and isochoric processes in Joule-Brayton [J-B] and Otto cycles respectively.

The results obtained here are intended largely for future statistical-mechanical and thermodynamic calculations. We hope to further to explore its thermodynamic applications.

## REFERENCES

1. Y. G. Rudoy and E. O. Oladimeji, Rudn J. Math. Inf. Sci. Phys. **25**, 276 (2017).

2. Y. G. Rudoy and E. O. Oladimeji, Phys. High. Educ. **23**, 20 (2017).

3. Y. G. Rudoy and E. O. Oladimeji, Phys. High. Educ. **23**, 11 (2017).

4. R. Kosloff, J. Chem. Phys. **80**, 1625 (1984).

5. T. Feldmann, Am. J. Phys. **64**, 485 (1996).

6. E. Geva and R. Kosloff, J. Chem. Phys. **96**, 3054 (1992).

7. A. Sisman and H. Saygin, J. Phys. D. Appl. Phys. **32**, 664 (1999).

8. R. U. and R. Kosloff, New J. Phys. **16**, 95003 (2014).

9. A. S. L. M. and A. J. S. and P. Kammerlander, New J. Phys. **17**, 45027 (2015).

10. M. B. and A. X. and A. F. and G. D. C. and N. K. and M. Paternostro, New J. Phys. **17**, 35016 (2015).

11. H. T. Quan, Y. Liu, C. P. Sun, and F. Nori, Phys. Rev. E **76**, 31105 (2007).

12. S. Whalen, M. Thompson, D. Bahr, C. Richards, and R. Richards, Sensors Actuators, A Phys. **104**, 290 (2003).

13. P. G. Steeneken, K. Le Phan, M. J. Goossens, G. E. J. Koops, G. J. A. M. Brom, C. Van Der Avoort, and J. T. M. Van Beek, Nat. Phys. **7**, 354 (2011).

14. J. P. Brantut, C. Grenier, J. Meineke, D. Stadler, S. Krinner, C. Kollath, T. Esslinger, and A. Georges, Science (80-. ). **342**, 713 (2013).

15. I. A. Martínez, É. Roldán, L. Dinis, D. Petrov, and J. M. R. Parrondo, **12**, 67 (2016).

16. T. Hugel, N. B. Holland, A. Cattani, L. Moroder, M. Seitz, and H. E. Gaub, Science (80-. ). **296**, 1103 (2002).

17. C. M. Bender, D. C. Brody, and B. K. Meister, J. Phys. A. Math. Gen. **33**, 4427 (2000).

18. F. Constantinescu, *Problems in Quantum Mechanics*, 1st ed. (Infosearch Limited, Moscow, 1972).

19. S. Flugge, *Practical Quantum Mechanics Volume II* (Springer, Berlin, 1971).

20. S.-H. H. Dong, *Factorization Method in Quantum Mechanics*, 1st ed. (Springer Netherlands, The Netherlands, 2007).

21. H. Hellmann, *Hans Hellmann: Einführung in Die Quantenchemie* (Springer Spektrum, Leipzig, 2015).

22. R. P. Feynman, Phys. Rev. **56**, 340 (1939).

23. L. J. Fernández-Alcázar, H. M. Pastawski, and R. A. Bustos-Marún, Phys. Rev. B **95**, (2017).

24. H. L. Calvo, F. D. Ribetto, and R. A. Bustos-Marún, Phys. Rev. B **96**, 1 (2017).

25. L. Guzmán-Vargas, V. Granados, and R. D. Mota, AIP Conf. Proc. **643**, 291 (2002).

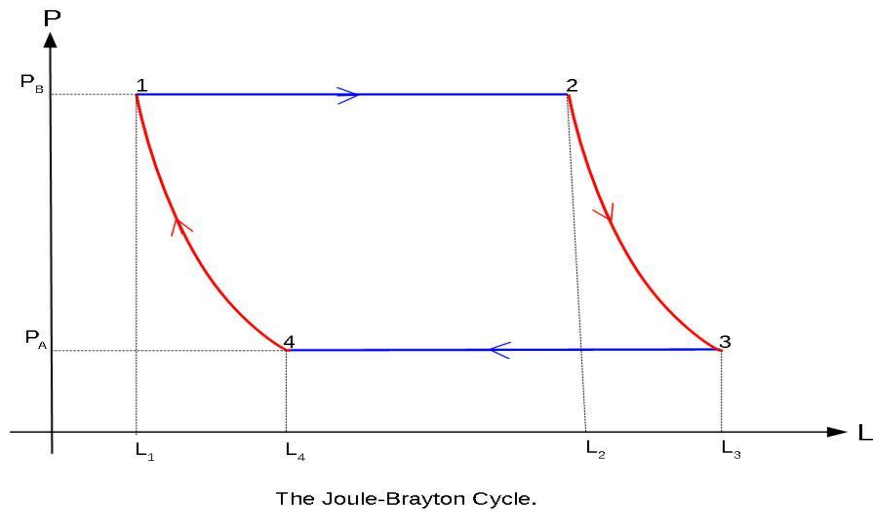

*Figure 1: the schematic representation of the Joule-Brayton's cycle.*

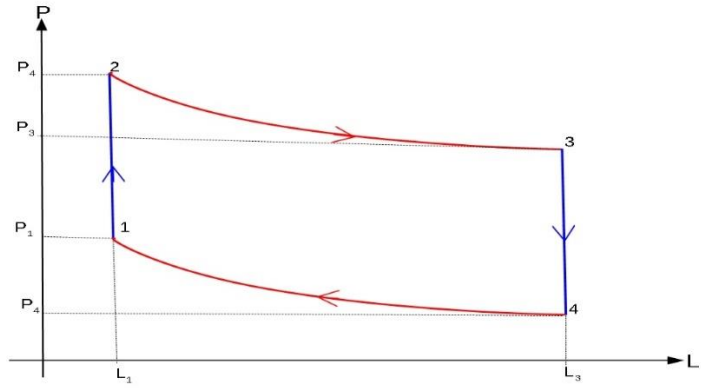

*Figure 2: the schematic representation of the Otto's cycle*